\documentstyle[12pt]{article}
\begin{document}
\title{Conformal Invariance and the exact solution of BFKL equations}
\author{
H. Navelet and R. Peschanski\\CEA, Service de Physique
Th\' eorique, CE-Saclay\\ F-91191 Gif-sur-Yvette Cedex, France}
\maketitle
\begin{abstract}
The conformal invariance properties of the QCD Pomeron in the transverse
plane allow us to give an explicit analytical expression for the solution of the BFKL equations both in the transverse coordinate and momentum spaces. This result is obtained from the solution of the conformal eigenvectors in the mixed representation in terms of two conformal blocks, each block being the product of an holomorphic times an antiholomorphic function.  
This property is used to give an exact expression for the QCD dipole multiplicities and dipole-dipole cross-sections in the whole parameter space, proving the equivalence between the BFKL and dipole representations of the QCD Pomeron.
\end{abstract}
\bigskip
{\bf 1.} Introduction
\bigskip

In his inspiring study \cite{lip} Lev Lipatov has shown that the equation obeyed by the BFKL kernel \cite{BFKL} of the   bare QCD Pomeron is invariant by the (global) conformal group of transformations in the tranverse coordinate space. Using a complete basis of conformal eigenfunctions $E^{n,\nu},$ he is able to express the elastic off-mass-shell gluon-gluon amplitude as an expansion over this basis, when $n,$ the conformal spin, is an  integer and $\nu$ corresponds to a continuous imaginary scaling dimension. In order to investigate the physical properties of this expansion, he is led to consider the eigenfunctions in a mixed representation, $E^{n,\nu}_q,$ which is obtained by a suitable Fourier transform 
of the $E^{n,\nu}.$ Given the nodal importance of these $q-$dependent eigenfunctions for the determination of the solutions of the BFKL equation and of their properties, it appears useful to go a step further. There already exists numerical \cite {salam} and various approximate analytic \cite {lip,muella,samuel} estimates of these quantities.  The aim of our paper
is to derive the exact analytical expressions for the elements of this basis. We perform this derivation by using a powerful method taking advantage of the conformal invariant properties of the theory. 

To be more specific, we find that the eigenfunctions  $E^{n,\nu}_q$ exhibit
 a structure of {\it conformal blocks} which already appeared, in particular, in the computation of correlation functions in 2-dimensional conformal-invariant quantum field theories\cite{correl}.  This structure is a finite linear combination of functions factorized in terms of holomorphic and antiholomorphic parts. The coefficients of this combination are such that they preserve the singlevaluedness of the physical quantities in the complex plane of the coordinates. This structure is then reproduced under different forms in all 
the relevant quantities in the QCD Pomeron calculations. 

In section 2 we give the expression of the $E^{n,\nu}_q$ in terms of their two conformal blocks. Then, in section 3, we derive the analytical expression in conformal blocks for 
$f_\omega (k,k',q)$ which can be interpreted \cite{lip} as the t-channel partial-wave amplitude for gluon-gluon scattering with gluon virtualities $-(k)^2,$ $-(k')^2,$ $-(q-k)^2,$ $-(q-k')^2;$ $t=-q^2$ is the momentum transfer and $\omega$ is the the Mellin-conjugate of the c.o.m. energy squared $s.$ In the next section 4 we give an exact expression of the dipole multiplicity which has been recently introduced and used in an asymptotic approximate form by Al Mueller \cite{muella}. It turns out that this quantity, which coincides with the Pomeron Green function recently calculated by Lipatov \cite{lipa}, obeys the conformal-block structure. In the last section 5, we determine 
the impact-parameter-dependent amplitude, first in the Lipatov original formulation and second in the Mueller approach and exhibit the non-trivial mathematical property which proves their identity in the whole parameter-space.

Our main results can be summarized as follows:

i) The exact expression of the BFKL amplitude for elastic dipole-dipole scattering at any fixed transverse momentum $q,$
see in the text formulae (\ref{11}, \ref{12}, \ref{13}), using the expression
(\ref{9})
found for $E^{n,\nu}_q$ .

ii) The  exact expression of the BFKL amplitude for gluon-gluon scattering in transverse momentum  space 
see in the text formulae (\ref{21}, \ref{22}), using the expression
(\ref{19})
found for the conformal blocks in the transverse momentum representation.

\bigskip
{\bf 2.} Conformal blocks for eigenfunctions
\bigskip

Using the notations of Ref. \cite{lip}, the eigenfunctions in coordinate space
are defined as:

\begin{equation}
E^{n,\nu}\left(\rho_{10},\rho_{20}\right) = (-1)^n \ \left(\frac {\rho_{10}\rho_{20}}{\rho_{12}}\right)^{\mu-1/2}
 \left(\frac {\bar \rho_{10}\bar \rho_{20}}{\bar \rho_{12}}\right)^{\tilde \mu-1/2} ,
\label{1}
\end{equation}
where $\rho_{ij}\equiv \rho_i-\rho_j$ and $\rho_i \ (i=0,1,2)$ are complex transverse coordinates. The conformal dimensions are defined as :
\begin{equation}
\mu = n/2 -i\nu\ ;\ \tilde \mu =-n/2 -i\nu,
\label{2}
\end{equation}
where $n,$ the conformal spin, is an integer. These are the eigenfunctions of the two Casimir operators of the conformal algebra, namely:
\begin{eqnarray}
\rho_{12}^2 \partial _1 \partial _2 E^{n,\nu}&=&\lambda _{n,\nu} E^{n,\nu} 
\nonumber \\
\bar \rho_{12}^2 \bar \partial _1 \bar \partial _2 E^{n,\nu}&=&\lambda _{n,-\nu} E^{n,\nu}
\nonumber \\
\lambda _{n,\nu} = 1/4 - \mu^2&,&\ \lambda _{n,-\nu} = \bar \lambda _{n,\nu}.
\label{3}
\end{eqnarray}

Lipatov introduces the following mixed representation of the eigenfunctions:
\begin{equation}
E^{n,\nu}_q (\rho) = \frac {2\pi^2}{b_{n,\nu}} \frac {1} {\vert\rho\vert}
\int {dz d\bar z\  e^{\frac {i}{2} (\bar q z + q \bar z)}\ 
E^{n,\nu}\left(\! z\! +\! \rho/2, z\! -\! \rho/2\right)}\ ,
\label{4}
\end{equation}
where
\begin{equation}
b_{n,\nu}=\frac {2^{4i\nu}\pi^3} {\vert n\vert/2 -i\nu} \
\frac {\Gamma (\vert n\vert/2\! -\!i\nu\!+\!1/2) \Gamma (\vert n\vert/2\! +\!i\nu)}
{\Gamma (\vert n\vert/2 \!+\!i\nu\!+\!1/2) \Gamma (\vert n\vert/2 \!-\!i\nu)}.
\end{equation}

Using the eigenvalue equations (\ref{3}), we derive the corresponding differential equations obeyed by the $E^{n,\nu}_q.$
\begin{eqnarray}
\left\{\frac {\partial^2}{\partial y^2} + \frac 1y \frac {\partial}{\partial y}
+\left(1-\frac {\mu^2}{y^2}\right)\right\} E^{n,\nu}_q (\rho) = 0,\nonumber \\
\left\{\frac {\partial^2}{\partial \bar y^2} + \frac 1 {\bar y} \frac {\partial}{\partial \bar y}
+\left(1-\frac {\tilde \mu^2}{\bar y^2}\right)\right\} E^{n,\nu}_q (\rho) = 0,\label{5}
\end{eqnarray}
where $y=\bar q \rho/ 4 .$  Each of these equations admits two linearly independant solutions which are $\bar q ^{\mp \mu} J_{\pm \mu} (y)$ and $ q ^{\mp \tilde \mu}J_{\pm \tilde \mu} (\bar y).$ So, the generic solution for  $E^{n,\nu}_q$ reads
\begin{equation}
 E^{n,\nu}_q (\rho) =  \bar q ^{-\mu} q ^{- \tilde \mu}\ \sum _{\alpha = \pm \mu,\beta = \pm \tilde \mu} C_{\alpha,\beta}\  J_\alpha (y)  J_\beta (\bar y),
\label{6}
\end{equation}
where $C_{\alpha,\beta}$ are constants. Now, the requirement 
 is that  the solution be a monovalued function with respect to the complex variable $\rho.$ This implies that only the combinations $(\mu,\tilde \mu)$ and 
$(-\mu,-\tilde \mu)$ contribute, in order to match the phases of the Bessel functions for each product. This finally yields:
\begin{equation}
 E^{n,\nu}_q (\rho) =  \bar q ^{-\mu} q ^{- \tilde \mu}\ \left[ C_{\mu, \tilde \mu}  J_{\mu} (y)  J_{\tilde \mu}(\bar y) +  C_{-\mu, -\tilde \mu}  J_{-\mu} (y)  J_{-\tilde \mu}(\bar y)\right].
\label{7}
\end{equation}
We determine the coefficients by matching with the known behaviour \cite {lip} when $\vert \rho q\vert \simeq 0,$ namely:
\begin{equation}
 E^{n,\nu}_q (\rho)\vert_{\rho q \rightarrow 0} =  \rho ^{-\mu} \bar \rho ^{- \tilde \mu}\ \left[\ 1 \ + e^{i\delta (n,\nu)}\  \left( \bar q \rho \right) ^{-2\mu} \left( q \bar \rho \right)^{- 2\tilde \mu}  \right],
\label{approx}
\end{equation}
where the phase may be written as:
\begin{equation}
e^{i\delta(n,\nu)}= 2^{- 12 i\nu} \ (-1)^{n+1} \ 
\frac {\Gamma (\vert n\vert/2\! -\!i\nu\!+\!1)\ \Gamma (-\vert n\vert/2\! -\!i\nu\!+\!1)}
{\Gamma (\vert n\vert/2 \!+\!i\nu\!+\!1) \ \Gamma (-\vert n\vert/2 \!+\!i\nu\!+\!1)}.
\label{phase}
\end{equation}
Using $J_{\sigma}(z)\vert_{z\rightarrow 0} = \left( z/ 2\right)^{\sigma} /\Gamma(\sigma + 1),$ we get
 
\begin{eqnarray}
 \frac {C_{\mu, \tilde \mu}} {  C_{-\mu, -\tilde \mu}} = (-1)^{n+1}\ 
\\  C_{-\mu, -\tilde \mu} = (-1)^{n+1} 2^{-6i\nu}  \Gamma \left(1-i\nu +\vert n \vert/2 \right)  \ \Gamma \left(1-i\nu - \vert n \vert/ 2 \right).
\label{8}
\end{eqnarray}
 
This finally yields:
\begin{eqnarray}
 E^{n,\nu}_q (\rho) =\bar q ^{i\nu \!-\! n/2} q ^{i\nu \!+\! n/2}\ 
 2^{-6i\nu} \Gamma \left(1\!-\!i\nu \!+\!\vert n \vert /2 \right) \Gamma \left(1\!-\!i\nu \!- \!\vert n \vert /2 \right)\times \nonumber \\\left[J_{n/2 -i\nu} (\frac {\bar q \rho} 4)  J_{-n/2-i\nu}(\frac {q \bar \rho} 4) - (-1)^n J_{-n/2 + i\nu} (\frac {\bar q \rho} 4)  J_{n/2 +i\nu}(\frac {q \bar \rho} 4)\right].
\label{9}
\end{eqnarray}
The obtained equation obviously satisfies the known relations\cite{lip} between eigenvectors:
\begin{equation}
\bar  E^{n,\nu}_q (\rho) \equiv E^{-n,-\nu}_q (\rho)= e^{-i\delta(n,\nu)} \vert
q\vert^{-4i\nu} \left(\frac{\bar q}q\right)^n  E^{n,\nu}_q (\rho).
\label{relation}
\end{equation}

The form  (\ref {9}) exhibits the behaviour of $E^{n,\nu}_q (\rho)$ for small 
$q \rho.$ An alternative form is more suitable to discuss the asymptotic
behaviour. Introducing the Hankel functions as follows\cite {gradstein}:
\begin{equation}
H_{\sigma}^{(1,2)} =  \frac{\Gamma (1/2 -\sigma)\ (y/2)^{\sigma}}{i\pi^{3/2}}
\int_{{\cal C}_{(1,2)}}\ e^{izt} \ (z^2-1)^{\sigma-1/2}\ dz,
\label{10}
\end{equation}
we replace $2 J_{\sigma} \equiv H_{\sigma}^{(1)} + H_{\sigma}^{(2)}$ in formula (\ref {9}) and get:
\begin{eqnarray}
 E^{n,\nu}_q (\rho) =\bar q ^{i\nu \!-\! n/2} q ^{i\nu \!+\! n/2}\ 
 2^{-6i\nu} \Gamma \left(1\!-\!i\nu \!+\!\vert n \vert /2 \right) \Gamma \left(1\!-\!i\nu \!- \!\vert n \vert /2 \right)\times \nonumber \\ \frac i2 sin(\mu\pi) \left[ H_{\mu}^{(2)} \left(\frac {\bar q \rho} 4 \right)  H_{\tilde\mu}^{(2)} \left(\frac {q \bar \rho} 4 \right) e^{-i\mu\pi} - H_{\mu}^{(1)}\left(\frac {\bar q \rho} 4 \right)  H_{\tilde\mu}^{(1)} \left(\frac {q \bar \rho} 4 \right) e^{i\mu\pi} \right].
\label{hankel}
\end{eqnarray}
Using this form, one easily gets the oscillatory asymptotic behaviour $$E^{n,\nu}_q (\rho) \propto  \frac 1 {\vert\rho \vert} \sin \left( {\cal R}e\left(\frac {\rho \bar q} 2\right) - \pi \frac n 2\right) $$ for large 
$ \rho.$ Note that the previous form (\ref {9}) implies  cancellations of the exponential
behaviour between the two terms. Conversely, cancellations are present in formula (\ref {hankel}) for small argument.

The form obtained in formulae (\ref {9}) and  (\ref {hankel}) is very reminiscent of   the one obtained with the conformal block structure of correlation functions in 2-dimensional conformal field theories \cite {correl}. It corresponds to the holomorphic factorization of the integrand in expression (\ref {4}).   The Hankel functions appearing in formula 
(\ref {hankel}) come from the  contour integral representations of the holomorphic and antiholomorphic factors of the integrand. However, it is known in conformal field theory that the actual result is a combination of conformal blocks whose coefficients contain contributions from the discontinuities of the integrand along the cuts. Strictly speaking, we cannot use in the present case the theorems applied in conformal field theories, due to the essential singularity corresponding to the Fourier transform. The theorems have been applied to solutions of the type of hypergeometric functions where the convergence conditions of the one-dimensional integrals in the complex plane are verified. Since we cannot rely on the known general theorems, we will propose
and use a different method applying directly to the 2-dimensional integrals. By a direct inspection of our results, we will show that the general theorems do apply. As an illustration of the method and a check of our result (\ref{9}) we show in Appendix {\bf A1} how to recover the $E^{n,\nu}$ from the inverse Fourier transform of $E^{n,\nu}_q,$ namely
\begin{equation}
E^{n,\nu}\left(\rho_{10},\rho_{20}\right)
 = \frac {b_{n,\nu}} {8\pi^4} \ \vert\rho_{12}\vert
\int d^2 q\  e^{-i \frac {q}{2} (\rho_{10}+\rho_{20})}\ E^{n,\nu}_q (\rho_{12}).
\label{inverse}
\end{equation}

\bigskip
{\bf 3.} Computation of the QCD Pomeron amplitudes
\bigskip

Formula (\ref {9}) gives directly the expression of the t-channel Pomeron partial wave amplitude $f_\omega ^q (\rho,\rho')$ in the mixed representation \cite {lip}. let us define  
\begin{equation}
f^{n,\nu}_q (\rho,\rho') = \frac {\vert\rho \rho'\vert} {16} \ \bar 
E^{n,\nu}_q (\rho')\ E^{n,\nu}_q (\rho) \left(\left[\nu^2\! +\!\left(\frac {n\!-\!1} 2\right)^2\right]
\left[\nu^2\! +\!\left(\frac {n\!+\!1} 2\right)^2\right]\right)^{-1},
\label{11}
\end{equation}
where 
\begin{equation}
f_\omega ^q (\rho,\rho')= \sum_{n= -\infty}^{+\infty}\ \int_{\nu = -\infty}^{+\infty}\ d\nu f^{n,\nu}_q (\rho,\rho')\ \frac {1}{\omega -\omega(\nu,n)},
\label{12}
\end{equation}
and \cite {lip}
\begin{equation}
\omega(\nu,n)= \frac {2 \alpha_S N_c} \pi \left(\Psi (1) - Re \left\{\Psi \left(\frac {\vert n\vert + 1 } 2 + i \nu \right)\right\}\right).
\label{13}
\end{equation}
Analogously, one can introduce the corresponding amplitudes in the impact-parameter space, namely $f_\omega (\rho_1,\rho_2,\rho_1',\rho_2')$ and 
$f^{n,\nu} (\rho_1,\rho_2,\rho_1',\rho_2').$ These functions are the Fourier-transforms of the previous ones (\ref {12},\ref {13}), namely:
\begin{eqnarray}
f^{n,\nu} (\rho_1,\rho_2,\rho_1',\rho_2') = \int\  d^2q \ e^{-iq \frac {\rho_{11'} +
\rho_{22'}} 2}\ \frac {\vert\rho \rho'\vert} {16}\ \bar E^{n,\nu}_q (\rho') \ E^{n,\nu}_q (\rho)\nonumber \\ \times \ \left(\left[\nu^2\! +\!\left(\frac {n\!-\!1} 2\right)^2\right]
\left[\nu^2 \!+\!\left(\frac {n\!+\!1} 2\right)^2\right]\right)^{-1}.
\label{14}
\end{eqnarray}
Finally, one defines the amplitudes in the momentum space by \cite {lip}
\begin{eqnarray}
\delta^2 (q-q') f^{n,\nu} (k,k',q) = (2\pi)^{-8} \int  d^2\rho_1 d^2\rho_2 d^2\rho_1' d^2\rho_2'\times \nonumber \\ \times e^{ik\rho_1 +i (k-q) \rho_2 -i k'\rho_1' -i(q'-k')\rho_2'}
f^{n,\nu} (\rho_1,\rho_2,\rho_1',\rho_2'),
\label{15}
\end{eqnarray}
where $k,$ $k',$ $q-k$ and $q'-k'$ are the transverse momenta of the external off-shell gluons (with propagators included). Strictly speaking, the integral defining $f^{n,\nu} (k,k',q)$ is divergent when  $\rho_i \rightarrow \infty.$ The actual physical amplitude is obtained by including  vertex functions\cite {lip} making convergent  the integral. Thus (\ref{15}) has to be understood as a distribution. 

Let us now derive the analytic solution for $f^{n,\nu} (k,k',q).$ 
For this sake, we introduce the new variables of integration:
\begin{equation}
\rho = \rho_1\!-\!\rho_2;\ \rho' = \rho_1'\!-\!\rho_2';\ b= \frac {\rho_{11'} \!+\!
\rho_{22'}} 2;\ \sigma = \frac {\rho_1\!+\!\rho_2\!+\!\rho_1'\!+\!\rho_2'} 2.
\label{16}
\end{equation}
The integration over $\sigma$ gives the expected $\delta^2 (q-q'). $ The result reads:
\begin{eqnarray}
f^{n,\nu} (k,k',q) = \pi (2\pi)^{-8} \int  d^2\rho d^2\rho' d^2b \ e^{i(k-q/2)\rho} e^{-i (k'-q/2)\rho'} e^{iqb} \times \ \nonumber \\
\times \ f^{n,\nu} (\rho_1,\rho_2,\rho_1',\rho_2').
\label{17}
\end{eqnarray}

With the definition (\ref {15}), the integration over $b$ gives:
\begin{eqnarray}
f^{n,\nu} (k,k',q) = \pi (2\pi)^{-6} 
 \left(\left[\nu^2\! +\!\left(\frac {n\!-\!1} 2\right)^2\right]
\left[\nu^2 \!+\!\left(\frac {n\!+\!1} 2\right)^2\right]\right)^{-1}\ 
\times
\nonumber \\
\ \times\ \int  d^2\rho \ e^{i(k-q/2)\rho}\  \frac {\vert\rho\vert} 4  \ E^{n,\nu}_q (\rho) \ \int  d^2\rho'\ e^{i(k'-q/2)\rho'}\  \frac {\vert\rho'\vert} 4 \ \bar E^{n,\nu}_q (\rho'),
\label{18}
\end{eqnarray}
which exhibits the factorization between the integrals over $\rho$ and $\rho'.$ 

As abovementionned the integrals in (\ref{18}) are divergent since the integrands behave as $\rho e^{ik\rho}$ ( resp. $\rho' e^{ik'\rho'}$). The computation can be made by regularizing the integrals by a factor $\rho^{\alpha}$ (resp.$\rho'^{\alpha}$), where $\alpha < -3/2$ can be formally driven to $0$ at the end of the calculation.

In order to compute
the integral say, over $\rho$ one again makes use of the conformal-block structure. Using expression (\ref{9}) for $ E^{n,\nu}_q (\rho),$ one is led to  conformal blocks of the form:
\begin{eqnarray}
B_{\pm}(\bar q,\bar k)= \int d\rho\ e^{\frac i 2 \rho \left(\bar k - \frac {\bar q} 2\right)}\ \rho^{1/2+\alpha/2}\ J_{\pm \mu} \left(\frac {\rho \bar q} 4\right)\ \ \ \ \ 
\nonumber \\ 
= \ \frac {\left(i\bar q/4\bar k\right)^{\pm \mu}}{\left( \bar k /2i \right)^{3/2+\alpha/2}}\ \  \ \frac {\Gamma(\pm \mu + 3/2 + \alpha/2)}{\Gamma(\pm \mu + 1)}\ \times \ \ \ \ \ \ 
\nonumber \\
2F_1 \left( \pm \mu +\! 3/2\!+\!\alpha/2,\pm \mu +1/2 
;\pm 2\mu + 1;\frac {\bar q}{\bar k}\right) 
.
\label{19}
\end{eqnarray}

The same equation holds for the quantities $\tilde B_{\pm} (q,k)$ by changing $(\mu,\bar q, \bar k)$ $\rightarrow $ $(\tilde \mu,q,k)$ in the previous expression. 

In Appendix {\bf A2}, we show how these conformal blocks appear in the computation of the 2-dimensional integrals of formula (\ref{18}). Taking into account the convergence properties ensured at large $\rho$ by our regularization procedure, one is led to only evaluate  the contributions of these integrals at small $\rho.$ One obtains  the following conformal block structure:
\begin{eqnarray}
I^{n,\nu}(k,q) \equiv \int \ d^2\rho \  e^{i(k-q/2)\rho} \ \frac {\vert\rho\vert} 4 \ E^{n,\nu}_q (\rho)  =  \nonumber \\ 
=2^{-6i\nu} q^{-\mu}\bar q ^{-\tilde\mu}\ \Gamma(1+\mu) \ \Gamma(1+\tilde\mu)\sin (\pi (\mu +1/2 + \alpha/2)) \nonumber\\ 
\ \times \ \left[B_+(\bar q,\bar k) \tilde B_+(q,k) - (-)^n \frac{\sin (\pi (-\mu +1/2 + \alpha/2))}
{ \sin (\pi (\mu +1/2 + \alpha/2))}
B_-(\bar q,\bar k) \tilde B_-(q,k)\right].
\label{21}
\end{eqnarray}
The final result for the amplitude $f^{n,\nu} (k,k',q)$ is
\begin{eqnarray}
f^{n,\nu} (k,k',q) &=& \left(\frac {1+(-)^n}2\right) \pi (2\pi)^{-6}\times\nonumber \\
&\times&  \left(\left[\nu^2 +\left(\frac {n-1} 2\right)^2\right]
\left[\nu^2 +\left(\frac {n+1} 2\right)^2\right]\right)^{-1} \ \times\nonumber \\
&\times&I^{n,\nu}(k,q)\ \bar I^{n,\nu}(k',q),
\label{22}
\end{eqnarray}
where we have taken $\alpha = 0$ by analytic continuation of formulae (\ref{19}-\ref{21}). Note that this amplitude is $0$ for odd values of the conformal spin $n.$

Notice the $sin$ coefficients in front of each conformal block in formula (\ref{19}) similar to those which  come from the discontinuities of the integrand along the cuts in actual computations of correlation functions\cite{correl}. As in the previous occurrence of conformal blocks, see section 2., the general theorems seem to be also valid  in the case where exist an essential singularity as in (\ref{18}).

\bigskip
{\bf 4.} Green functions and QCD dipole multiplicities
\bigskip

In the QCD dipole formalism \cite {muellb, muella}, one defines  the multiplicity of dipoles of (2-dimensional) transverse size $\rho'$ originated from an initial dipole of transverse size $\rho$ in an high-energy onium-onium scattering with a transverse momentum $q$. In terms of the $(n,\nu)$ representation, it reads \begin{equation}
N^{n,\nu}_q = 
 \frac {\vert\rho\vert} {\vert\rho'\vert}\  E^{n,\nu}_q (\rho) \ \bar E^{n,\nu}_q (\rho').
\label{23}
\end{equation}
 The expression for the $ E^{n,\nu}_q$ corresponding to formula (\ref{9}) gives an explicit analytic realization of the dipole multiplicity in the whole phase space.

A  Fourier transform leads to  the multiplicity of dipoles at an impact parameter $b.$ In terms of the $(n,\nu)$ representation, one writes \cite {muella}
\begin{equation}
N^{n,\nu} (\rho,\rho',b) = 
\int  \ dq  d\bar q\  e^{i/2(q\bar b + \bar q b)} \ \frac {\vert\rho\vert} {\vert\rho'\vert}\  E^{n,\nu}_q (\rho) \ \bar E^{n,\nu}_q (\rho').
\label{24}
\end{equation}
 Using again the expression for the $ E^{n,\nu}_q$, see (\ref{9}), we get the following conformal blocks (for the holomorphic part):
\begin{eqnarray}
A_{\pm}(\rho,\rho',b)&=& \int d\bar q\ e^{\frac i 2 {\bar q} b} \ J_{\pm\mu }\left(\frac {\rho \bar q} 4\right)\ J_{\pm\mu} \left(\frac {\rho' \bar q} 4\right)
\nonumber \\
&=& Q_{\pm\mu\!-\!1/2} \left(\frac {b^2\!-\!\left(\frac {\rho} 2 \right)^2\! -\!\left(\frac {\rho'} 2 \right)^2}{2 \left(\frac {\rho} 2 \right) \left(\frac {\rho'} 2 \right)}\right),
\label{25}
\end{eqnarray}
where $ Q$ is the Legendre function of second kind \cite{gradstein}. The antiholomorphic blocks $\tilde A_{\pm} $ are obtained by changing $\mu \rightarrow \tilde \mu, z \rightarrow \bar z.$

In Appendix {\bf A3}, using the same method of 2-dimensional integration, we find the following combination of conformal blocks: 
\begin{eqnarray}
N^{n,\nu} (\rho,\rho',b) =\frac {16} {\pi^2 \vert\rho'\vert^2} \ \times \ \ \ \ \ \ \ \ \ \ \ \ \ \ \ \ 
\nonumber\\
\ \times \ \Gamma(\!-n/2\!+\!i\nu\!+\!1)\ \Gamma(\!-n/2\!-\!i\nu\!+\!1) \Gamma(\!n/2\!+\!i\nu\!+\!1)\ \Gamma(\!n/2\!-\!i\nu\!+\!1)\ \nonumber\\
 \times \ \sin (2\pi\mu)\left[A_+(z) \tilde A_+(\bar z) - A_-(z) \tilde A_-(\bar z)\right],
\label{26}
\end{eqnarray}
where
\begin{equation} 
z \equiv \left(\frac {b^2\!-\!\left(\frac {\rho} 2 \right)^2\! -\!\left(\frac {\rho'} 2 \right)^2}{2 \left(\frac {\rho} 2 \right) \left(\frac {\rho'} 2 \right)}\right).
\label{27}
\end{equation}

As a matter of fact, up to a normalization factor this multiplicity is nothing but the gluon Green function recently calculated by Lev Lipatov \cite {lipa}. Indeed, using the following change of variables:
\begin{equation}
x \equiv \frac {\rho_{11'} \rho_{22'}} { \rho_{12} \rho_{1'2'}} = \frac {1-z} 2; \ h=n/2\!+\!i\nu\!+\!1/2;\ \tilde h = \!-\!n/2 \!+\!i\nu\!+\!1/2,
\label{28}
\end{equation}
where $x$ is the well-known complex anharmonic ratio, and the known \cite{gradstein} definition of the Legendre functions in terms of $ _2F_1,$ one recovers the Lipatov result. By definition the Green function reads
\begin{eqnarray}
G_{n,\nu} \equiv \int \ d^2\rho_0 \bar E^{n,\nu}\left(\rho_{1'0},\rho_{2'0}\right) E^{n,\nu}\left(\rho_{10},\rho_{20}\right)
\nonumber \\
=\frac 1 {\pi^4} \vert b_{n,\nu}\vert^2 \vert \rho \rho'\vert \int \ d^2q\ e^{-iq \frac {\rho_{11'} +
\rho_{22'}} 2}  \ \bar E^{n,\nu}_q (\rho') E^{n,\nu}_q (\rho),
\label{29}
\end{eqnarray}
where $\vert b_{n,\nu}\vert^2  = \frac {\pi^6} {\nu^2 + n^2/4}.$ One finally obtains the identity valid in the whole phase space:
\begin{equation}
N^{n,\nu} (\rho,\rho',b) \equiv  \frac {\nu^2 + n^2/4} {\pi^2\vert \rho'\vert^2} \ G_{n,\nu}. 
\label{30}
\end{equation}
The final equation of this section establishes the exact equivalence between the BFKL elastic amplitudes with the QCD dipole multiplicities.

\bigskip
{\bf 5.} Dipole-dipole elastic amplitude
\bigskip

In the BFKL formalism the function $f^{n,\nu} (\rho_1,\rho_2,\rho_1',\rho_2'),
$ see (\ref {15}), can be interpreted as the dipole-dipole elastic amplitude
where  dipoles of size $\rho_{12}$ and $\rho_{1'2'}$ collide at  an impact parameter distance $b= \frac {\rho_{11'} + \rho_{22'}} 2.$ It is straightforward to get:
\begin{equation}
f^{n,\nu} (\rho_1,\rho_2,\rho_1',\rho_2') \equiv \ \left(\nu^2 + n^2/4\right) G_{n,\nu} = \pi^2 \vert \rho'\vert ^2
N^{n,\nu} (\rho,\rho',b) .
\label{31}
\end{equation}

In the Mueller formalism \cite{muella}, the same amplitude between the two incoming  dipoles is evaluated in a different way. The two dipoles interact through the cascading of dipoles with decreasing c.o.m.rapidity. In the central region the  dipoles obtained after cascading at a given impact-parameter interact through the elementary gluon-gluon exchange at $q=0.$ However, this formulation destroys the conformal invariance of the formalism. In terms of a fully conformal-invariant formalism one  writes:
\begin{eqnarray}
f^{n,\nu} (\rho_1,\rho_2,\rho_1',\rho_2') = \sum_{m,m'} \ \int \ \frac {d^2\sigma_{12} d^2\sigma_{1'2'}} {\vert \sigma_{12}\ \sigma_{1'2'}\vert^2}
\ d^2c\ d^2c' \ \delta (b-c-c')
\nonumber \\
\int d\mu\
N^{m,\mu} (\rho_{12},\sigma_{12},c)
\int d\mu'\
N^{m',\mu'} (\rho_{1'2'},\sigma_{1'2'},c')\ 
f^{n,\nu}_q \left(\sigma_{12},\sigma_{1'2'}\right),
\label{32}
\end{eqnarray}
where one considers more generally the $q \ne 0$ amplitude. This equation can
easily be expressed in terms of the dipole multiplicities with a transverse momentum $q,$ see (\ref{23}), namely
\begin{eqnarray}
f^{n,\nu} (\rho_1,\rho_2,\rho_1',\rho_2') = \sum_{m,m'} \ \int \ \frac {d^2\sigma_{12} d^2\sigma_{1'2'}} {\vert \sigma_{12}\ \sigma_{1'2'}\vert^2}
 \ d^2q \  e^{iqb}
\nonumber \\ \times
\int d\mu\
N_q^{m,\mu} (\rho_{12},\sigma_{12})
\int d\mu'\
N_q^{m',\mu'} (\rho_{1'2'},\sigma_{1'2'})\ 
\frac {\vert \sigma_{1'2'} \vert ^2} {16} 
\nonumber \\ \times
\left\{\left[\nu^2 +\left(\frac {n-1} 2\right)^2\right]
\left[\nu^2 +\left(\frac {n+1} 2\right)^2\right]\right\}^{-1}
N_q^{n,\nu} \left(\sigma_{12},\sigma_{1'2'}\right),
\label{Nqexpression}
\end{eqnarray}

Noting the factorised form of the functions $N_q,$ see formula (\ref{23}),
the integration over $ \sigma_{12}, \sigma_{1'2'}$ can be easily performed by using the orthogonality properties of the eigenvectors $E_q.$ In Appendix {\bf A4} we derive these othogonality relations which read:
\begin{eqnarray}
\frac 1 {4\pi^2} \int\  \frac {d^2 \rho} {\vert \rho \vert^2} \ \
E^{n,\nu}_q (\rho)\  \bar E^{m,\mu}_q (\rho)\ \ = \ \nonumber \\  \ =
\left[\delta_{n,m}\ \delta(\nu-\mu) + \delta_{-n,m}\ \delta(\nu+\mu)
\vert q\vert^{4i\nu} \left(\frac q {\bar q}\right)^n e^{i\delta(n,\nu)}\right].
\label{orthogonal}
\end{eqnarray}
Then, the summations over $(m,\mu),$ (resp.  $(m',\mu')$ in formula (\ref{Nqexpression}) yield
\begin{eqnarray}
\sum_m  \int d\mu \ E^{m,\mu}_q (\rho)
\left[\delta_{n,m}\ \delta(\nu-\mu) + \delta_{-n,m}\ \delta(\nu+\mu)
\vert q\vert^{4i\nu} \left(\frac q {\bar q}\right)^n e^{i\delta(n,\nu)}\right]= \nonumber \\   =
E^{n,\nu}_q (\rho) + E^{-n,-\nu}_q (\rho)
\vert q\vert^{4i\nu} \left(\frac q {\bar q}\right)^n e^{i\delta(n,\nu)} \equiv 2 E^{n,\nu}_q (\rho),
\label{muintegral}
\end{eqnarray}
using the relation between $E^{n,\nu}_q$ and  $E^{-n,-\nu}_q,$ expression (\ref{relation}). The antiholomorphic part, with $(m',\mu'),$ similarly gives $
2 \bar E^{n,\nu}_q (\rho').$

Finally, the remaining $q$ integration formula (\ref {Nqexpression}) gives again relation (\ref{31}) and allows us  to  prove the identity between BFKL and dipole-dipole amplitudes. Interestingly enough, the restriction of the amplitudes to their value at $q=0,$ as calculated in\cite{muella} ignore the contributions coming from the $ E^{-n,-\nu}_q$ in formula (\ref{muintegral}). This leads to a factor 4 in the normalisation, due to the final result of (\ref{muintegral}). One finally get back to the relation (\ref{31}).

\bigskip
{\bf 6.} Conclusion and outlook
\bigskip

Using the conformal invariant properties of the BFKL Pomeron, we have been able to get exact analytic formulae for various quantities of interest in this formulation. The main result we obtain is an exact formulation of the off-shell gluon-gluon amplitude in the physical momentum space, thanks to the expression for the eigenvectors of the BFKL equation in the mixed representation as a sum of two conformal blocks implying Bessel functions of complex index and argument. We also obtain an exact expression for the QCD dipole multiplicities and scattering amplitudes, valid in the whole phase-space, proving the full equivalence between the BFKL and QCD dipole formulation for these quantities.   As a check we recover the recent result on the gluon Green function obtained in a different way by Lipatov \cite {lipa}.

The fact that exact expressions based on conformal invariance have been obtained for the eigenvectors in the mixed representation, is promising for handling the multi-Pomeron interactions \cite{muella,salam,gavin}.

\bigskip

{\bf Acknowledgments}
Discussions
with Michel Bauer, Jeff geronimo, Riccardo Guida and Samuel Wallon have been appreciated.

\eject
 {\bf APPENDICES}
 
\bigskip
{\bf Appendix A1.} Inverse Fourier transform of $ E^{n,\nu}_q (\rho).$
\bigskip

Up to constant factors in equation (\ref{9}), one has to evaluate the integral
\begin{eqnarray}
{\cal I} = \int q dq\ d\phi   e^{- \frac i{2} q R \cos\phi}
 \left( q e^{i\phi}\right) ^{-\mu} \left(q e^{-i\phi}\right)^{- \tilde \mu}\nonumber \\
\left[  J_{\mu} (y)  J_{\tilde \mu}(\bar y) - (-)^n   J_{-\mu} (y)  J_{-\tilde \mu}(\bar y)\right],
\label{Integral}
\end{eqnarray}
with $y=\frac {\bar q \rho}4 = \vert q\rho\vert e^{i\theta},$
and , in the two-dimensional integration plane, $\theta = \psi - \phi$ is the angle $\hat \rho \hat q,$
$\psi = \hat\rho \hat R $ and $\phi = \hat q \hat R.$

The integrand in formula (\ref{Integral}) behaves as $\vert q\vert^{-2(\mu + \tilde \mu) +1}$ when $q \rightarrow 0$ and as $\vert q\vert^{-2(\mu + \tilde \mu) -1/2}$ when $q \rightarrow \infty$ (after the angular integration).  Thus the inverse fourier transform can be defined by analytic continuation of formula (\ref{Integral}) from the region   $1/4<\mu + \tilde \mu<1.$  Using  these convergence properties the method for computing ${\cal I}$  is to take only in consideration the contribution at the origin. We thus expand the Bessel functions
\begin{equation}
 J_{\sigma} (y) = \left(\frac y2\right)^{\sigma} \sum_k \frac {(-)^k\
\left(\frac y2\right)^{2k}}{k!\ \Gamma(\sigma+1+k)}
\label{Bessel}
\end{equation}
and use the following identities:
\begin{eqnarray}
\frac1{2\pi} \int_0^{2\pi}\ e^{i(qb\cos\theta - m\theta)}\ d\theta\ = J_m (qb)
\nonumber \\
\int_0^N\ q^{\beta}\ dq\ J_m(qb) = (-)^m\ \frac 1{2\pi} \left[\frac 2 b\right]^{\beta + 1}\ \frac {\Gamma\left(\frac{1+\beta+\vert m\vert}2\right)}
{\Gamma\left(\frac{1-\beta-\vert m\vert}2\right)} + f(N),
\label{identities}
\end{eqnarray}
for $m$ integer.
The final result for $\cal I$ does not depend on the upper bound contribution $f(N)$
due to the abovementionned convergence properties.

In formula (\ref{Integral}), the term $J_{\mu} J_{\tilde\mu}$ corresponds to 
$\beta = 1+2k +2\tilde k$ and $m=2k - 2\tilde k$ in the identities 
(\ref{identities}). This contribution is $0$, since $\Gamma\left(\frac{1-\beta-\vert m\vert}2\right)^{-1}=\Gamma(-2k)^{-1}\equiv 0,$
if $k \le \tilde k,$ or $\Gamma(-2\tilde k)^{-1}\equiv 0$ if $k > \tilde k.$
the term $J_{-\mu} J_{-\tilde\mu}$ corresponds to 
$\beta = 1+2k +2\tilde k-2(\mu + \tilde \mu)$ and $m=2k - 2\tilde k-2(\mu -\tilde \mu).$ 

Using the same identities 
(\ref{identities}), we get factorisation in $k$ and  $\tilde k,$ with coefficients 
\begin{eqnarray}
\sin 2\pi\tilde \mu \ \Gamma(1+2k-2\mu) \Gamma(1+2\tilde k-2\tilde \mu)
\ {\rm if}\  k>\tilde k,
\nonumber \\
\sin 2\pi\mu \ \Gamma(1+2k-2\mu) \Gamma(1+2\tilde k-2\tilde \mu)
\ {\rm if}\  k\le\tilde k,
\label{factors}
\end{eqnarray}
which are indeed identical since $\mu -\tilde \mu=-n$ is an integer. Using straightforward doubling formulae, we recover  formula (\ref{1}).

\bigskip
{\bf Appendix A2.} Calculation  of $ \int  d^2\rho \ e^{iv\rho}\  \vert\rho\vert^{\alpha}  \ E^{n,\nu}_q (\rho).$
\bigskip

Up to constant factors, one has to evaluate the integral
\begin{eqnarray}
{\cal J} = \int  d\rho d\phi   e^{i q v \cos\phi}
\rho^{1+\alpha} \times 
\nonumber \\
\times \left[  J_{\mu} (y)  J_{\tilde \mu}(\bar y) - (-)^n   J_{-\mu} (y)  J_{-\tilde \mu}(\bar y)\right],
\label{Jntegral}
\end{eqnarray}
with $y=  q\rho e^{i\theta},$
and , in the two-dimensional integration plane, $\theta = \psi - \phi$ is the angle $\hat \rho \hat q,$
$\phi = \hat\rho \hat v $ and $\psi = \hat q \hat v.$

The integrand in formula (\ref{Jntegral}) behaves as $\rho^{1+\beta+2(\mu + \tilde \mu)}$ when $\rho \rightarrow 0$ and as $\rho^{1+\alpha-3/2}$ when $\rho\rightarrow \infty$ (after the angular integration).  Thus the integral can be defined by analytic continuation of formula (\ref{Integral}) from the region   $-2<\alpha<-1/2$ (for imaginary $\mu + \tilde \mu$).  Using  these convergence properties the method for computing ${\cal J}$  is again to take only in consideration the contribution at the origin.

In formula (\ref{Jntegral}), the term $J_{\mu} J_{\tilde\mu}$ corresponds to 
$\beta = \alpha +1+2k +2\tilde k +\mu +\tilde \mu$ and $m=2k - 2\tilde k +\mu -\tilde \mu$ in the identities 
(\ref{identities}). Using the same identities (\ref{identities})
 we get factorisation in $k$ and  $\tilde k,$ with coefficients 
\begin{eqnarray}
\sin \pi (\tilde \mu + \alpha/2) \Gamma(\alpha/2 +1+2k+\mu) \Gamma(\alpha/2 +1+2\tilde k +\tilde \mu)
\ {\rm if}\  k>\tilde k,
\nonumber \\
\sin \pi ( \mu + \alpha/2) \Gamma(\alpha/2 +1+2k+\tilde \mu) \Gamma(\alpha/2 +1+2\tilde k + \mu)
\ {\rm if}\  k\le\tilde k,
\label{Jfactors}
\end{eqnarray}
which are indeed identical if $\mu -\tilde \mu=-n$ is an even integer, and opposite for an odd one. Using straightforward doubling formulae, we recover  formula (\ref{21}).

\bigskip
{\bf Appendix A3.} Calculation  of the dipole multiplicity $N^{n,\nu} (\rho,\rho',b).$
\bigskip

Up to constant factors, one has to evaluate the integral
\begin{eqnarray}
{\cal K} = \int  q dq d\phi   e^{i q b \cos\phi} \times \nonumber
\\ \times \left[  J_{\mu} (y)  J_{\tilde \mu}(\bar y) - (-)^n   J_{-\mu} (y)  J_{-\tilde \mu}(\bar y)\right] \times \nonumber \\
\times  \left[  J_{-\mu} (y')  J_{-\tilde \mu}(\bar y') - (-)^n   J_{\mu} (y')  J_{\tilde\mu}(\bar y')\right]
,
\label{Kntegral}
\end{eqnarray}
with $ y=  q\rho e^{i\theta},\ y'=  q\rho' e^{i\theta'},$
and, in the two-dimensional integration plane, $\theta = \psi - \phi$ is the angle $\hat \rho \hat q,$  $\theta' = \psi' - \phi$ is the angle $\hat \rho' \hat q,$
$\psi = \hat\rho \hat b $ and $\psi' = \hat\rho' \hat b.$

The integrand in formula (\ref{Kntegral}) behaves as $q^{-3/2}$ when $q\rightarrow \infty$ (after the angular integration) and is regular at $q\rightarrow 0.$  Thus the integral is well defined. The method for computing ${\cal K}$  is again to take only in consideration the contribution at the origin.

In formula (\ref{Kntegral}), the term $J_{\mu} (y)  J_{\tilde \mu}(\bar y)
J_{-\mu} (y')  J_{-\tilde \mu}(\bar y')$ corresponds to 
$\beta = 1+2k +2\tilde k + 2k' +2\tilde k'$ and $m=2k - 2\tilde k+2k' - 2\tilde k'$ in the identities 
(\ref{identities}). This contribution is $0$, since $\Gamma\left(\frac{1-\beta-\vert m\vert}2\right)^{-1}=\Gamma(-2k-2k')^{-1}\equiv 0,$
if $k+k' \le \tilde k+\tilde k',$ or $\Gamma(-2\tilde k-2\tilde k')^{-1}\equiv 0$ if $k+k' > \tilde k+\tilde k'.$ The same for the combination 
$J_{-\mu} (y)  J_{-\tilde \mu}(\bar y)
J_{\mu} (y')  J_{\tilde \mu}(\bar y').$
Note that this is consistent with the requirement of monovaluedness which is not satisfied by the these two first factors. The other terms give a non-zero contribution.
the term $(-)^{n+1} J_{\mu} (y)  J_{\tilde \mu}(\bar y) J_{-\mu} (y')  J_{-\tilde \mu}(\bar y')$
 corresponds to 
$\beta = 1+2k+2k' +2\tilde k+2\tilde k'+2(\mu + \tilde \mu)$ and $m=2k+2k' - 2\tilde k-2\tilde k'+2(\mu -\tilde \mu).$ Using the same identities 
(\ref{identities}), we get factorisation in $k,k'$ and resp. $\tilde k,\tilde k',$ with coefficients 
\begin{eqnarray}
\sin 2\pi\tilde \mu \ \Gamma(1+2k+2k'+2\mu) \Gamma(1+2\tilde k+2\tilde k'-2\tilde \mu)
\ {\rm if}\  k+k'>\tilde k+\tilde k,
\nonumber \\
\sin 2\pi\mu \ \Gamma(1+2k+2k'+2\mu) \Gamma(1+2\tilde k+2\tilde k'-2\tilde \mu)
\ {\rm if}\  k+k'\le\tilde k+\tilde k'
,
\label{factorsprime}
\end{eqnarray}
which are indeed identical since $\mu -\tilde \mu=-n$ is an integer.
 Using straightforward doubling formulae, the sums over $k,k',$ (resp.$\tilde k,2\tilde k',$  lead formally to $F_4$ generalized hypergeometric functions of
Appel and Kampe et Ferri\'e\cite{gradstein} with complex conjugate arguments.
We thus recover again the conformal block structure. By inspection of their
parameters, the  $F_4$ functions reduce to ordinary hypergeometric functions (or Legendre functions) leading to formula  (\ref{24}) after adding the other contribution with $n\rightarrow -n,$ $\nu \rightarrow -\nu.$ 

\bigskip
{\bf Appendix A4.} Orthogonality relations of $E^{n,\nu}_q (\rho).$

Up to constant factors, one has to evaluate the integral
\begin{equation}
{\cal L} = \int  \rho^{-1} d\rho d\phi   \ \left[ E^{n,\nu}_q (\rho)\  \bar E^{m,\mu}_q (\rho)\ \right].
\label{Lntegral}
\end{equation}

The integrand in formula (\ref{Lntegral}) behaves as $\rho^{-7/2}$ when $\rho\rightarrow \infty$ (after the angular integration). Using  these convergence properties the method for computing ${\cal L}$  is again to take only in consideration the contributions at the origin which are given by:
\begin{equation}
 E^{n,\nu}_q (\rho)\vert_{\rho q \rightarrow 0} =  \rho ^{-\mu} \bar \rho ^{- \tilde \mu}\ \left[\ 1 \ + e^{i\delta (n,\nu)}\  \left( \bar q \rho \right) ^{-2\mu} \left( q \bar \rho \right)^{- 2\tilde \mu}  \right].
\label{approxim}
\end{equation}
Using straightforwardly the identities $\int_0^{2\pi}\ d\phi\ e^{im\phi}\equiv 2\pi\delta_{m,0}$ and $\int \ d\rho \rho^{2i\sigma-1}\equiv \pi\delta (\sigma),$
we get formula (\ref{orthogonal}). Note that the same relation can be obtained from the orthogonal properties\cite{lip} of the eigenvectors $E^{n,\nu}\left(\rho_{10},\rho_{20}\right)$ by Fourier transform.

.

\eject


\begin{thebibliography}{99}
\bibitem{lip}
L.N. Lipatov {\it Zh. Eksp. Teor. Fiz.}{\bf 90} (1986) 1536 (Eng. trans.
{\it Sov. Phys. JETP} {\bf 63} (1986) 904.

\bibitem{BFKL}
L.N. Lipatov, {\it Sov. J. Nucl. Phys.} {\bf 23} (1976) 642;
V.S. Fadin, E.A. Kuraev and L.N. Lipatov, {\it Phys. lett.} {\bf B60} (1975)
50;
E.A. Kuraev, L.N. Lipatov and V.S. Fadin, {\it Sov.Phys.JETP} {\bf 44} (1976) 45, {\bf 45} (1977) 199; 
I.I. Balitsky and L.N. Lipatov, {\it Sov.J.Nucl.Phys.} {\bf 28} (1978) 822

\bibitem{salam}
G.~P. Salam, {\it Nucl. Phys.} {\bf B449} (1995) 589; {\bf B461} (1996) 512;


\bibitem{muella}
A.H.Mueller, {\it Nucl. Phys.} {\bf B437} (199) 107.

\bibitem{samuel}
H. Navelet and S. Wallon, to appear.

\bibitem{correl}
A.M. Polyakov, {\it Zh. Eksp. Teor. Fiz.} { Lett.\bf 12} (1970) 538, {\bf 66} (1974) 23;
A.A. Migdal, {\it Phys. lett.} {\bf B44} (1972) 112;
A.A. Belavin, A.M. Polyakov and A.B. Zamolodchikov,  {\it Nucl. Phys.} {\bf B241} (1984) 333; 
Vl.S. Dotsenko and V.A. Fateev, 
{\it Nucl. Phys.} {\bf B240} FS{\bf12} (1984) 312; 

For a collection of  reprints on conformal invariance, see
{\it Conformal invariance and applications to statistical mechanics}, C. Itzykson, H. Saleur, J.-B. Zuber eds., World Scientific, 1988.
\bibitem{lipa}
L.N. Lipatov {\it Small-x physics in perturbative QCD}, hep-ph/9610276.

\bibitem{gradstein}
I.S. Gradshteyn and I.M. Ryzhik {\it Table of integrals and products}
(Academic Press, Inc., A. Jeffrey ed., 1994)

\bibitem{muellb}
A.H.Mueller, {\it Nucl. Phys.} {\bf B415} (1994) 373; 
A.H.Mueller and B.Patel, {\it Nucl. Phys.} {\bf B425} (1994) 471.

\bibitem{guida}
R. Guida, private communication; R. Guida and N. Magnoli {\it  Tricritical Ising model near criticality}, to appear; S.D. Mathur, {\it Nucl. Phys.} {\bf B369} (1992) 433.

\bibitem{gavin}
A.~H. Mueller and
 G.~P. Salam, {\it Nucl. Phys.} {\bf B475} (1996) 293;
A.~Bialas and  R.~Peschanski, {\it Phys. lett.} {\bf B378} (1996) 302;  {\bf B387} (1996) 405.


\end{thebibliography}
\end{document}